\documentclass[preprint]{revtex4}
\usepackage{epsfig}
\usepackage{graphicx}
\usepackage{dcolumn}
\usepackage{bm}
\def\la{\lower.5ex\hbox{$\; \buildrel < \over \sim \;$}}
\def\ga{\lower.5ex\hbox{$\; \buildrel > \over \sim \;$}}
\def\apj{ApJ}
\def\mnras{MNRAS}
\def\prd{PRD}

\def\physrep{Physics Reports}

\begin{document}


\title{Tangled magnetic fields and CMBR signal from reionization epoch}
\author{Rajesh Gopal}\email{gopal@rri.res.in}\author{Shiv K. Sethi}\email{sethi@rri.res.in}
\affiliation{Raman Research Institute, Bangalore 560080, India}


\begin{abstract}
We compute the secondary CMBR anisotropy signal from the reionization of 
the universe in the presence of tangled magnetic fields. We consider 
the tangled-magnetic-field-induced 
scalar, vector, and tensor modes for our analysis. 
The most interesting signal for $\ell \la 100$ arise from tensor perturbations. In particular, we show that the enhancement 
observed by WMAP  
 in the TE cross-correlation signal for $\ell \la 10$    could be explained 
by tensor TE cross-correlation from tangled magnetic fields
generated during the inflationary epoch   for 
magnetic field strength $B_0 \simeq 4.5 \times 10^{-9} \, \rm G$ and magnetic
field power spectrum spectral index $n \simeq -2.9$. Alternatively,
a mixture of tensor mode signal with primordial scalar modes give weaker
bounds on the value of the optical depth to the reionization surface, 
$\tau_{\rm reion}$: $\tau_{\rm reion} = 0.11 \pm 0.02$.  This 
analysis can also 
 be translated 
to a limit on magnetic field strength of $\simeq 5 \times 10^{-9} \, \rm G$
for wave-numbers $\la 0.05 \, \rm Mpc^{-1}$. 
\end{abstract}

\maketitle

\section{Introduction}

 Coherent magnetic field of micro-gauss strength 
are observed in galaxies and clusters of galaxies (\cite{parker},\cite{zeldovich}, for a recent review 
see e.g. \cite{widrow}).  Observational evidence exist for even larger
scale magnetic fields \cite{kim}. 
The  origin of these observed magnetic fields however   is not well
understood.
The observed magnetic fields
could have arisen from dynamo amplification of small seed 
($\la 10^{-20} \, \rm G$)  magnetic fields (see e.g. \cite{ruzmaikin}; \cite{shukurov})
which originated from various astrophysical processes in the early universe 
(\cite{harrison}, \cite{subramanian1}, \cite{kulsrud}, \cite{grasso}, 
 \cite{widrow}, \cite{gopal1}, \cite{matarrese}) . Alternatively
the magnetic fields of nano-Gauss strength could have originated from 
some early universe process like electroweak phase transition or 
during  inflation (e.g. \cite{turner},  \cite{ratra}, see \cite{grasso},
\cite{giovannini} for reviews). In this scenario, 
the observed micro-gauss 
magnetic fields  then result from adiabatic compression of this primordial
magnetic field.

The existence of primordial magnetic fields of nano-Gauss strength can
influence the large scale structure formation in the universe (\cite{wasserman}, \cite{kim1}, \cite{subramanian2}, 
\cite{sethi}, \cite{gopal}, \cite{sethi1}). Also 
these magnetic fields could leave observable signatures in the CMBR
anisotropies (\cite{barrow}, \cite{subramanian3}, \cite{subramanian4}, \cite{durrer}, \cite{seshadri}, 
\cite{mack}, \cite{lewis}).

In recent years, the study of
CMBR anisotropies has proved to be the best probe of 
the theories of structure formation in the universe (see e.g. \cite{hu1} for 
a recent review).  The simplest model of 
scalar, adiabatic perturbations, generated during the inflationary 
era, appear to be in good agreement with both the CMBR anisotropy measurements 
and the distribution of matter at the present epoch (see e.g. \cite{spergel},
\cite{tegmark}). Tensor perturbations 
could have been sourced by primordial gravitational waves during the 
inflationary epoch. There is no definitive evidence of the existence  of 
tensor perturbations in the CMBR anisotropy data;  the  
 WMAP experiment, from temperature anisotropy data,  obtained
upper limits on the amplitude of tensor perturbations \cite{spergel}.  
Vector perturbations
are generally not considered in the standard analysis 
 as the primordial vector perturbations would have 
decayed by the epoch of recombination in the absence of a source.
 An indisputable signal of vector 
and tensor modes is that unlike scalar modes 
 these perturbations generate $B$-type CMBR polarization anisotropies (see 
e.g. \cite{hu4} and references therein). 
At present, only upper limits exist on this polarization mode \cite{kovac}.
 However, the 
on-going CMBR 
probe WMAP and the upcoming experiment  Planck surveyor
 have the capability of unravelling 
the effects of vector and tensor perturbations.

Recent WMAP results suggest that the universe underwent an 
epoch of re-ionization at  $z \simeq 15$; in particular WMAP analysis
concluded that the optical depth to the last reionization surface is $\tau_{\rm reion} = 0.17 \pm 0.04$ \cite{kogut}; which  means that nearly 20\% of CMBR
 photons re-scattered during the period  of reionization. The secondary 
anisotropies
 generated 
during this re-scattering leave interesting signatures  especially in 
CMBR polarization anisotropies (see e.g. \cite{zaldarriaga}), as is
evidenced by the recent WMAP results \cite{kogut}. 

Primordial magnetic fields source all  three kinds of perturbations. 
In this paper we study 
the secondary CMBR 
anisotropies, generated during the epoch of reionization,
 from vector, tensor, and scalar  modes,  in the 
presence of primordial tangled magnetic fields. Recently, Lewis \cite{lewis}
computed fully-numerically CMBR vector and tensor
 temperature and polarization anisotropies in the presence of magnetic fields
including the effects of reionization. Seshadri and Subramanian 
 \cite{seshadri1}
calculated the secondary temperature  anisotropies from
vector modes  owing to reionization.  Our approach is to compute 
the secondary temperature and polarization anisotropies 
semi-analytically by identifying the main sources of anisotropies in each
case; we compute the anisotropies by   using the formalism of \cite{hu4}.
We also compute the tensor primary signal  to compare with the 
already existing analytical results  for  tensor
anisotropies \cite{mack} . 

In the next section, we set up the preliminaries by discussing the 
models for primordial magnetic fields and the process of reionization. 
In \S 3, \S 4, and  \S 5, we consider vector, tensor, and scalar modes.  In \S 6  the detectability of the signal is discussed.
 In \S 7, we present and 
summarize our conclusions.  While presenting numerical results in this 
paper, we use 
 the currently-favoured FRW  model: spatially flat
with $\Omega_m = 0.3$ and $\Omega_\Lambda = 0.7$ (\cite{spergel}, \cite{perlmutter}, \cite{riess}) with  $\Omega_b h^2 = 0.024$ (\cite{spergel}, \cite{tytler}) and 
$h = 0.7$ (\cite{freedman}).

\section{Primordial magnetic fields, reionization, and CMBR anisotropies}

Assuming that the tangled magnetic fields are generated by some process
in the early universe, e.g. during inflationary epoch,
 magnetic fields at large
scales ($\ga 0.1 \, \rm Mpc$) are
  not affected appreciably by different processes 
in either the pre-recombination or the post-recombination universe
(\cite{sethi1}, \cite{jedamzik}, \cite{subramanian2}). 
In this  regime,  the magnetic field decays
 as $1/a^2$ from the expansion of the universe. This allows us to express: 
${\bf B}({\bf x},\eta)=\tilde{\bf B}({\bf x})/a^{2}$; here ${\bf x}$ is the comoving coordinate. We further assume  tangled magnetic fields,
$\tilde{\bf B}$, present in the early universe, 
 to be  an isotropic,  homogeneous, and Gaussian 
 random process.
This allows one to write, 
in Fourier space (see .e.g. \cite{landau}):
\begin{equation}
\langle {\tilde{B}_i({\bf q})\tilde{B}^*_j({\bf k})} \rangle 
= \delta^3_{\scriptscriptstyle D}({\bf q -k}) 
\left (\delta_{ij} - k_i k_j/k^2 \right ) M(k)
\label{eq:n6}
\end{equation}
Here $M(k)$ is the magnetic field power spectrum and 
$k=\vert {\bf k}\vert$ is the comoving wavenumber. We assume 
a power-law magnetic field power spectrum here: $M(k) = Ak^n$. We 
consider the range of scales between $k_{\rm min}$ taken to be zero here
 and  small scale  cut-off at $k=k_{\rm max}$; $k_{\rm max}$ is 
determined by the effects of damping by radiative viscosity
before recombination. Following Jedamzik et~al. \cite{jedamzik},
 $k_{\rm max} \simeq 60 \, {\rm Mpc^{-1}} (B_0/(3 \times 10^{-9} \, \rm G)$;
 $B_0$ is   the   RMS of magnetic field fluctuations  at the present epoch.  
 $A$ can be calculated  by fixing  the value of 
the RMS of the magnetic field, $B_0$,  smoothed at a given scale, $k_{\rm c}$.
 Using a sharp $k$-space filter, we get, 
 \begin{equation}
A = {\pi^2 (3+n) \over k_c^{(3+n)}} B_0^2 
\label{normal}
\end{equation}
We take $k_c = 1 \rm Mpc^{-1}$ throughout this paper. For $n \simeq -3$, 
the spectral indices of interest in this paper,
the  RMS filtered at  any scale has weak dependence on the scale of filtering.

Recent WMAP observations showed that the universe might have got ionized 
at redshifts $z \simeq 15$. However the details of the ionization history 
of the universe during the reionization era are still unknown; for instance
the universe might have got reionized at $z = 15$ and remained fully ionized 
till the present or the universe might have got partially reionized 
with ionized fraction $x_e \la 0.3$  at $z \simeq 30$ and became fully
 ionized for $z \la 10$. Both these ionization histories are compatible with 
the WMAP results \cite{kogut}. Given this lack of knowledge we model the 
reionization history by assuming the following visibility function, which 
gives the normalized probability that 
the photon last scattered between epoch $\eta$ and $\eta+d\eta$, to model
the period of reionization:
\begin{equation}
g(\eta,\eta_0) \equiv \dot \tau \exp(-\tau) = {(1-\exp(-\tau_{\rm reion})) \over \sqrt{\pi} \Delta\eta_{\rm reion}}\exp\left[-(\eta-\eta_{\rm reion})^2/\Delta\eta_{\rm reion}^2 \right ]
\label{eq:visfun}
\end{equation}
Here $\tau(\eta,\eta_0) = \int_{\eta_0}^\eta n_e\sigma_t dt$ is 
the optical depth from Thompson scattering; 
$\tau_{\rm reion}$ is the optical depth to the epoch of reionization;
for compatibility with WMAP results, we use $\tau_{\rm reion} = 0.17$ 
throughout. $\eta_{\rm reion}$ and $\Delta\eta_{\rm reion}$ are the epoch of 
reionization and the width of reionization phase, respectively; we take $\eta_{\rm reion}$ corresponding to $z_{\rm reion} = 15$ and $\Delta\eta_{\rm reion} = 0.25 \eta_{\rm reion}$. Notice that the visibility function is normalized
to $\tau_{\rm reion}$ for $\tau_{\rm reion} << 1$.

\section{CMBR anisotropies from vector modes}

From a given wave number $k$ of vector perturbations, the contribution 
to CMBR temperature 
and polarization anisotropies to  a given angular mode $\ell$
  can be expressed as (see e.g. \cite{hu4}):
\begin{eqnarray}
{\Theta_{T\ell}^{\rm v}(\eta_0,k) \over (2 \ell +1)} & = & \int_0^{\eta_0} d\eta \, \exp(-\tau)\left [\dot \tau(v_{\rm b}^{\rm v} - V) j_\ell^{(11)}[k(\eta_0 - \eta)] + (\dot\tau P^{\rm v}(\eta) + {1\over \sqrt{3}} kV)j_\ell^{(21)}[k(\eta_o -\eta)] \right ] \label{temp}\\
{\Theta_{E\ell}^{\rm v}(\eta_0,k) \over (2 \ell +1)} & = & -\sqrt{6} \int_0^{\eta_0} d\eta \,\exp(-\tau)\dot \tau P^{\rm v}(\eta) \epsilon_\ell^{\rm v}[k(\eta_0 -\eta)] \label{emode}\\
{\Theta_{B\ell}^{\rm v}(\eta_0,k) \over  (2 \ell +1)}& = & -\sqrt{6} \int_0^{\eta_0} d\eta \,\exp(-\tau)\dot \tau P^{\rm v}(\eta) \beta_\ell^{\rm v}[k(\eta_0 -\eta)] \label{bmode}
\end{eqnarray}   
Here $v_{\rm b}^{\rm v}$  and $V$ are the line-of-sight  components
 of the vortical component
 of the baryon velocity and the vector metric perturbation. 
$P^{\rm v}(\eta) = {1/10} [\Theta_{T2}^{\rm v} - \sqrt{6} \Theta_{E2}^{\rm v} ]$ and 
the Bessel functions, $j_\ell$, $\epsilon_\ell$ and~$\beta_\ell$
 that give radial projection for a given mode are 
given in reference \cite{hu4}. The evolution of 
 vector metric perturbations, $V_i({\bf k},\eta)$ is determined from Einstein's equations (e.g. \cite{hu4}, \cite{mack}):
\begin{eqnarray}
\dot V_i  +  2 {\dot a  \over a} V_i & =  & -{16\pi G a^2  S_i({\bf k},\eta) \over k} \label{ein1}\\
-k^2 V_i & =  & 16 \pi G a^2 \sum_j (\rho_j + p_j)(v_{ij}^{\rm v} -V_i) \label{ein2}
\end{eqnarray} 
Here $S_i$, the source of vector perturbations,  is determined by primordial 
tangled magnetic field in this paper. The index $j$ correspond to baryonic, photons and dark matter vortical 
component of velocities. For tangled magnetic fields,  the vortical 
velocity component of the dark matter doesn't couple  to the source
of vector perturbations to linear order and $\Omega_i = v_i^{\rm v} - V_i$ decays as $1/a$ 
for dark matter (see e.g. \cite{mack}); and hence the dark matter contribution 
can  be dropped from the Einstein's equations. The photons couple 
to baryons through Thompson scattering. In the pre-recombination epoch, 
the photons 
are tightly coupled to the baryons as the time scale of Thompson scattering 
is short as compared to  the expansion rate; besides the photon 
density is comparable to baryon density at the epoch of recombination.
In the reionized models we consider here, neither the photons are 
tightly coupled to baryons nor are they dynamically important. Therefore 
photon contribution can also be dropped from  Eq.~(\ref{ein2}). 
 Eq.~(\ref{ein2}) then simplifies to:
\begin{equation}
-k^2 V_i  =   16 \pi G a^2 \rho_{\scriptscriptstyle B} \Omega_B^{\rm v} 
 \label{ein3}
 \end{equation}
with $\Omega_B^v = (v_{\rm b}^{\rm v} -V_i)$.
The quantity of interest is the angular power spectrum of the CMBR anisotropies
which is obtained from squaring Eqs.~(\ref{temp}),~(\ref{emode}), and~(\ref{bmode}),  taking ensemble average, and integrating over all $k$:
\begin{equation}
C_{\ell T,E,B} = {4 \over \pi}\int dk k^2 \left [{\Theta_{T, E,B\ell}(k,\eta_0) \over 2\ell +1} \right ]^2
\end{equation}
This expression is valid for both vector and tensor perturbations; for 
scalar perturbation the prefactor is $2/\pi$.

For primordial magnetic field, the sources $S_i({\bf k},\eta)$ of vector
perturbation (Eq.~(\ref{ein1})) is the vortical component of the Lorentz 
force:
\begin{equation}
S_i({\bf k},\eta) ={1 \over a^4 4 \pi} \hat k {\rm x} {\rm F.T.}[{\bf \tilde B}({\bf x}) {\rm x} (\nabla {\rm x} {\bf  \tilde B}({\bf x}))] \equiv S_i({\bf k}){1\over a^4}
\end{equation}
It can be  checked that this Newtonian expression for 
 $S_i$ is the same as the more rigorously defined
$\Pi_{i}^{V}$ in Appendix~A (Eq.~(\ref{vecso})).

\subsection{Temperature anisotropies from vector modes}
As seen from Eq.~(\ref{temp}), there are three sources of 
temperature anisotropies. The most important contribution comes  from 
 vorticity $\Omega_B^{\rm v}$. For the reionized models,  using Eqs.~(\ref{ein1})~and~(\ref{ein2}), it can be expressed as:
\begin{equation}
\Omega_B^{\rm v}({\bf k},\eta) = {k S_i({\bf k}) \eta \over a \rho_{b0}}
\label{vort}
\end{equation} 
 Here $\rho_{b0}$ is the baryon density at the present epoch. 
The other  major  contribution is from temperature quadrupole
 $\Theta_{T2}^{\rm v}$. 
For reionized models, the quadrupole at the epoch of reionization is 
dominated by the free-streaming of the dipole from the last scattering surface
(see discussion below, Eq.~(\ref{eq:quadr})). 
This contribution  is generally small 
but in this case can be comparable to the 
vorticity effects at small values of $\ell$. This is owing to the 
fact that the vorticity is decaying  and therefore during 
 reionization  epoch 
its contribution is smaller as compared
to  the epoch of recombination. The quadrupole term on the 
other hand gets its contribution from the vorticity computed 
at the epoch of recombination (Eq.~(\ref{eq:quadr})).
 This, as we shall discuss below, is
not the case for scalar and tensor anisotropies, as the dominant 
source  of anisotropy  is either constant (metric perturbations 
for tensor perturbations) or is increasing (compressional velocity mode
for scalar perturbation) as the universe evolves.  The third source 
of temperature anisotropies is  metric vector perturbation $V$;
this term can be 
 comparable to the other terms only at super-horizon scales. We drop this 
term in this paper. 

In Figure~1 we show the secondary temperature anisotropies generated 
during  the epoch of 
 reionization from vector modes.
 It is seen that the quadrupole term has significant 
contribution only for $\ell \la 20$. The dominant contribution 
at larger $\ell$  is 
from the vorticity during reionization.  The vorticity source contribution 
can be approximated as:
\begin{equation}
 {\Theta_{T\ell}^{\rm v}(\eta_0,k) \over (2 \ell +1)} \simeq {S(k) \over 4 \pi \rho_{\rm b0}} {\eta_0^2 \over \eta_{\rm reion}} j_\ell^{(11)}[k(\eta_0 - \eta_{\rm reion})] \tau_{\rm reion}
\end{equation}
for $\ell \le 20$ and 
 \begin{equation}
 {\Theta_{T\ell}^{\rm v}(\eta_0,k) \over (2 \ell +1)} \simeq {S(k) \over k}{1 \over  8 \pi \rho_{\rm b0}} {\eta_0^2  g(\eta_0-\ell/k,\eta_0) \over \eta_{\rm reion}} \sqrt{{\pi \over \ell}}
\end{equation} 
for $\ell \ga 50$. 
The temperature angular power spectrum 
 from vorticity increases roughly as $\ell^{2.4}$ for $\ell \ga 50$,
with the signal  reaching a value roughly 
$0.3 \, \rm \mu k$ at  $\ell \simeq 10^4$. This is in agreement with
the results of \cite{seshadri1}.

\subsection{Polarization  anisotropies from vector modes}
The main source of the polarization anisotropies is the temperature quadrupole
$\Theta_{T2}^{\rm v}$. One contribution to the temperature  quadrupole
at the epoch of reionization is from the free-streaming of the dipole 
from the last scattering surface. The dipole at the last scattering surface
can be obtained from the tight-coupling solutions 
to the temperature anisotropies \cite{mack}. The quadrupole from the 
free-streaming
 of the dipole at the epoch of recombination is:
\begin{equation}
\Theta_{T2}^{\rm v}({\bf k},\eta) =   5\Omega_B^{\rm v}({\bf k},\eta_{\rm rec}) j_2^{(11)}(k(\eta - \eta_{\rm rec}))
\label{eq:quadr}
\end{equation}
Here $\eta_{\rm rec}$ corresponds to the epoch of recombination. As is
 the case for scalar perturbation-induced polarization in the 
reionized model (e.g. \cite{zaldarriaga}) this quadrupole doesn't suffer the 
suppression as the quadrupole 
  prior to  the epoch of recombination when the photons and 
baryons are tightly coupled. The structure of anisotropies generated by 
the quadrupole is determined by $j_2^{(11)}(k\eta)$ around the epoch of 
reionization. This typically gives a peak in anisotropies at $\ell \simeq 2 \eta_0/\eta_{\rm reion}$. This source dominates the contribution to 
polarization anisotropies for $\ell \la 10$. Another contribution to the 
temperature quadrupole at the epoch of reionization comes from the secondary
temperature anisotropies generated at the epoch of reionization. The 
approximate value of this quadrupole can be got from retaining the 
first term in Eq.~(\ref{temp}):
\begin{equation}
{\Theta_{T2}^{\rm v}(\eta,k) \over (2 \ell +1)}  =  \int_0^{\eta} d\eta' \, g(\eta,\eta') \Omega_B^{\rm v}(\eta') j_2^{(11)}[k(\eta - \eta')]
\label{eq:quadr1}
\end{equation}
This contribution is generically smaller than the first contribution. 
Firstly, this depends on the vorticity evaluated close to the 
epoch of reionization as opposed to the first contribution which 
is proportional to the vorticity at the epoch of recombination. As the 
vorticity decays as $a^{-1/2}$ in the matter-dominated era (Eq.~(\ref{vort})),
the latter contribution is suppressed by nearly a factor a $100$ in the 
angular power spectrum. Second, as only a small fraction of photons re-scatter
(nearly 20\%), this contribution is further suppressed by a factor of 
$\tau_{\rm reion}^2$. However, this contribution is not suppressed at small
angular scales and, therefore,  might dominate 
the polarization anisotropies at large 
values of $\ell$. 

In Figure~2, we show the $E$ and $B$ polarization angular power spectrum
from the sources given by Eqs.~(\ref{eq:quadr})~and~(\ref{eq:quadr1}). 
As discussed above, the secondary polarization anisotropies are dominated 
by the quadrupole generated by free-streaming of dipole at the last
scattering surface. As expected for vector modes (\cite{hu4}), the $B$-mode signal
is larger than the $E$-mode  signal; the signal strength reaches $\simeq 
10^{-3} \, \rm \mu k$ at $\ell \simeq 10$ in both cases. This dominates 
the primary signal for $\ell \la 10$ as also seen in the 
numerical results  of   Lewis \cite{lewis}. The 
contribution from the quadrupole generated at the epoch of reionization is
seen to be  completely sub-dominant

\section{CMBR anisotropies from tensor modes}
The energy-momentum tensor for magnetic fields has a non-vanishing traceless,
transverse component which sources the corresponding tensor 
 metric perturbation. 
This in turn affects the propagation of radiation from the last scattering 
surface to the present and hence gets manifested as additional anisotropies.
 In this section we calculate the effect of reionization on the
 resultant anisotropies. For the temperature anisotropies, we study 
this effect, by calculating the power spectra separately for the standard 
recombination (no-reionization) and reionized scenario whereas for the 
polarization anisotropies we compute  the secondary anisotropies 
by using the visibility function given by Eq~(\ref{eq:visfun}).

\subsection{Tensor temperature anisotropies}

The line-of-sight integral solution for temperature anisotropies,
for tensor perturbations is given by (\cite{hu4}):
\begin{equation}
\frac{\Theta_{\ell T}^{T}(k,\eta_0)}{2\ell+1}=\int_{0}^{\eta_{0}}d\eta e^{-\tau}[\dot{\tau}P^{(T)}-\dot{h}]j_{\ell}^{(22)}[k(\eta_{0}-\eta)]
\label{tensigt}
\end{equation}
Here, $P^{T}(\eta) = {1/10} [\Theta_{T2}^{\rm T} - \sqrt{6} \Theta_{E2}^{\rm T} ]$ is the tensor polarization source and $\dot{h}$ is the
 gravitational wave contribution whose evolution is detailed in 
Appendix~B. The polarization source is modulated by the visibility function 
and hence is localized to the last-scattering surface. In the tight 
coupling limit before recombination, $P^T \simeq -\dot h/(3 \dot \tau)$ (\cite{mack}); for a more detailed derivation of $P^T$ in the tight-coupling regime
see Appendix~B.
In the post-recombination epoch,  $P^T$ is determined 
by the free-streaming of quadrupole generated at the last scattering surface. 
However, the visibility function is very small at  epochs prior to
reionization. Therefore 
the main contribution of this term comes only from epochs prior to 
recombination. The gravitational 
wave source on the other hand being modulated by the cumulative visibility 
$\exp({-\tau})$ contributes at all epochs.  
As a result, the $P^T$  contributes negligibly to temperature 
anisotropies at all multipoles for the case of standard recombination.  
In the reionized model, this term gets additional 
contribution from epochs close to  reionization redshift but continues to
be sub-dominant to the other term.  
 We have also checked this numerically.  Hence we can neglect the first
 term in the above solution and using the matter-dominated solution 
for $\dot{h}$ (Appendix~B)
 we arrive at the following expression for the angular power
spectrum:
\begin{equation}
C_{\ell T}^{T}=\frac{4}{\pi}\left(\frac{9R_{\gamma}}{\rho_{\gamma}}\right)^{2}\left(\frac{8(l+2)!}{3(l-2)!}\right)\int dk k^{2} \Pi_{T}^{2}(k)\left(\int_{x_{d}}^{x_{0}}dx \exp(-\tau)\frac{j_{2}(x)}{x}\frac{j_{l}(x_{0}-x)}{(x_{0}-x)^2} \right)^{2}
\label{tenpow}
\end{equation}
Here, $x \equiv k\eta$, $x_{0}\equiv k\eta_{0}$ and $x_{d}\equiv k\eta_{\rm rec}$.
The above expression is evaluated numerically for the two different 
ionization histories: standard recombination with and without reionization 
which are essentially characterized by the different behaviour of the 
cumulative visibility $\exp({-\tau})$. The temperature power
spectra are shown in Figure~3. As seen in the figure, the temperature
 power spectrum in both cases shows similar behaviour. The power is nearly
 flat upto $\ell \simeq 100$ after which the amplitude falls rapidly. 
This behaviour is identical to that obtained for primordial gravitational
 waves. This is expected because, the tensor metric perturbation is
 sourced by the magnetic field only upto the neutrino-decoupling epoch 
thereby imprinting an initial nearly scale-invariant spectrum after which
 the evolution is source-free. The effect of reionization is to reduce
 the cumulative visibility between recombination ($z \simeq 1100$) and
 reionization ($z \simeq  15$) epochs.
 This is why the signal  is suppressed for the reionized model.
  Approximate analytic expressions to primary $C_{\ell T}^{T}$ were 
derived in (\cite{mack}). However these give the correct qualitative behaviour
 $C_{\ell T}^T\propto \ell^{0.2}$ only for $\ell \la 100$. This is because 
in their analytic results,
 the lower limit for the time-integral is taken to be zero in
Eq.~(\ref{tenpow}) whereas
 the correct lower limit is $\eta_{\rm rec}$ since the cumulative visibility is
 zero for $\eta \la \eta_{\rm rec}$. We have not neglected this lower
 limit in our 
numerical calculation and hence we obtain  the damping behaviour
 for $\ell \ga 100$; Lewis \cite{lewis} also obtains this damping behaviour.
Our results are in reasonable agreement with the results of \cite{lewis} 
in the entire range of $\ell$;    these results 
also agree to within factors with the results 
 of \cite{mack} for $\ell  \la 75$ when the different 
convention we use for defining $B_0$ is taken into account. 
 Our results are quantitatively accurate to better
than $10\%$ for the lower
 multipoles $\ell \la 75$ but begin to differ appreciably 
from the results of numerical studies for larger $\ell$
or in the damping regime \cite{ng}.  This 
is because we have not treated the transition regime from 
radiation-dominated to matter-dominated for the gravitational wave 
evolution accurately. As described in Appendix~B, we have assumed 
instantaneous transition. This however does not affect the qualitative 
description of modes whose wavelength is greater than the
 transition-width $k \la \eta_{\rm eq}^{-1}$ which in turn corresponds to 
multipoles $\ell \la 300$.  

\subsection{Polarization anisotropies from tensor modes}
The line-of-sight solution for the E and B-mode polarization is given as:
\begin{equation}
\frac{\Theta_{\ell B}^T(k,\eta_{0})}{2l+1}=-\sqrt{6}\int_{0}^{\eta_{0}}d\eta  \dot\tau \exp({-\tau}) P^{T}\beta_{l}^{T}[k(\eta_{0}-\eta)]
\end{equation}
\begin{equation}
\frac{\Theta_{\ell E}^T(k,\eta_{0})}{2l+1}=-\sqrt{6}\int_{0}^{\eta_{0}}d\eta  \dot\tau \exp({-\tau}) P^{T}\epsilon_{l}^{T}[k(\eta_{0}-\eta)] \label{emodesig}
\end{equation}
Here, $\beta_l^{T}$ and $\epsilon_l^{T}$ are the tensor polarization
 radial functions as given in Ref.~\cite{hu4}. 
The tensor polarization source $P^{T}(\eta)$ in this case will contribute 
significantly to the above integral only close to the reionization epoch.
 There are two contributions to the polarization at $\eta_{\rm reion}$: 
one due to
 the quadrupole generated at the reionization surface and the other due to 
the free-streaming primary quadrupole. However, as in the case of 
vector perturbations,  the free-streaming primary
 quadrupole  will give the dominant contribution. We thus have,
\begin{equation}
P^{T}(k,\eta)=\frac{1}{10}\Theta_{T 2}^{T}(k,\eta)=-\frac{1}{2}\int_{\eta_{\rm rec}}^{\eta}d\eta \dot{h}j_2^{(22)}[k(\eta_{0}-\eta)] \exp(-\tau)
\label{tenqu}
\end{equation}
To simplify the calculations we make the following approximation. Since the
 visibility function is strongly peaked at $\eta_{\rm reion}$, we take $P^{T}$ 
outside the integral by evaluating it at the visibility 
peak $\eta_{\rm reion}$.
We have verified that this approximation works extremely well for the 
lower multipoles where the power is significant. We thus get the 
following expressions for the polarization angular power spectra:
\begin{equation}
C_{\ell B}^T=\frac{6}{\pi}\int dk k^{2}\Pi_{T}^{2}(k)\left[P^{T}(\eta_{\rm reion})\right]^{2}\left(\int_{\eta_{\rm rec}}^{\eta_{0}} d\eta \dot{\tau}e^{-\tau} \beta_{l}^{T}[k(\eta_{0}-\eta)]\right)^{2}  
\label{tenbmode}
\end{equation}
\begin{equation}
 C_{\ell E}^T=\frac{6}{\pi}\int dk k^{2}\Pi_{T}^{2}(k)\left[P^{T}(\eta_{\rm reion})\right]^{2}\left(\int_{\eta_{\rm rec}}^{\eta_{0}} d\eta \dot{\tau}e^{-\tau} \epsilon_{l}^{T}[k(\eta_{0}-\eta)]\right)^{2}  
\label{tenemode}
\end{equation}
As seen in the above expressions, the polarization power spectrum  is 
modulated by the visibility function itself instead of the cumulative 
visibility in the case of temperature power spectrum. As a result, both 
$E$ as well as $B$ mode anisotropies peak close to the multipole 
corresponding to the horizon scale at reionization. Physically this can be 
understood as follows: the modes which are super-horizon at reionization
 experience negligible integrated Sachs-Wolfe 
 effect before $\eta_{\rm reion}$ and 
hence very small 
polarization is generated for such modes. Maximum polarization is 
generated for modes that just enter the horizon at $\eta_{\rm reion}$. For 
sub-horizon  modes,  the amplitude of the gravitational wave falls 
 and then sets itself  into oscillations which is reflected as a drop 
in power for higher multipoles.

The polarization power spectra are shown in 
Figure~4. As seen in the figure, the E-mode power peaks 
at $\ell \sim 8$ whereas the B-mode power peaks at $\ell\sim 7$. The 
corresponding signal strengths at the peaks are $\sim 0.2\mu K$ 
in both cases. As expected, 
 the E-mode power
 is marginally greater than the B-mode power mainly because of the slightly 
different behaviour of the radial projection factors \cite{hu4}. The primary 
anisotropies for both the polarization modes is sub-dominant on these
 scales. This  enhancement in the net 
(primary+secondary) signal   was also seen in the numerical 
 calculations of Ref.~\cite{lewis}.

We also show the primary CMBR polarization  anisotropies from tensor modes
in Figure~4.
For computing these anisotropies we use the tight-coupling quadrupole, $P^T(\eta)$ as 
derived in Appendix~B (Eq.~(\ref{tenquadr})). The primary power spectra
are also computed  from Eqs.~(\ref{tenbmode}) and~(\ref{tenemode}) with
lower limit of the time integral replaced by zero. 
Our results are in agreement with the 
numerical results of \cite{lewis} when we take into account the fact that
we use different value of $\eta_\star/\eta_{\rm in}$ (Eq.~(\ref{finsol})): we use 
$\eta_\star/\eta_{\rm in}  = 10^{18}$, which gives the epoch of 
generation of the tangled magnetic field close to inflationary epoch. While 
presenting numerical results, Lewis \cite{lewis} uses $\eta_\star/\eta_{\rm in} = 10^{6}$, which puts the epoch of generation of magnetic field close to the 
epoch of electro-weak phase transition. Therefore our signal is roughly
an order of magnitude larger than the results of \cite{lewis}. 

In Figure~5 we show the expected TE cross-correlation from tensor modes,
computed using Eqs.~(\ref{tensigt}), (\ref{emodesig}), and~(\ref{tenqu}),  
including the effect of reionization. 
 The effect of reionization is seen as the peak in the 
TE cross-correlation for $\ell \la 10$. The signal is dominated by 
the primary signal for large multipoles. Note that the TE cross-correlation
is positive in the entire range  $\ell \la 150$ as was also pointed 
out by Ref.~\cite{mack} for the primary tensor TE cross-correlation (for
details of sign of TE cross-correlation for various modes see \cite{hu4}). 
In the next section we compare this signal with 
the WMAP observation of TE cross-correlation.

\section{Secondary CMBR  anisotropies from scalar modes}
In addition to the vortical component of the velocity field, the tangled
magnetic fields also generate  compressional  velocity 
fields which seed density perturbations. These density perturbations 
have interesting consequences for the formation of structures in the 
universe (\cite{wasserman}, \cite{kim1}, \cite{subramanian2}, 
\cite{sethi}, \cite{gopal}, \cite{sethi1}). The compressional velocity 
field also give rise to secondary anisotropies during the epoch of 
reionization. We compute this anisotropy here. The line-of-sight 
solution to the temperature anisotropies from these velocity perturbations is:
\begin{equation}
\frac{\Theta_\ell^{S}(k,\eta_0)}{2\ell+1}=\int_{0}^{\eta_{0}}d\eta e^{-\tau}\dot{\tau} v_{\rm b}^{\rm s}(k,\eta) j_{\ell}^{(10)}[k(\eta_{0}-\eta)]
\end{equation}
Here $v_{\rm b}^{\rm s}$ is the line-of-sight component of the compressional
velocity field. $j^{(10)}$ is defined in Ref.~\cite{hu4}.
 The growing mode of compressional velocity can be 
expressed as (\cite{wasserman}, \cite{gopal}):
\begin{equation}
v_{\rm b}^{\rm s}({\bf k},\eta) ={\eta \over 4 \pi \rho_{m0}} \hat k {\rm .} \left ( {\rm F.T.}[{\bf \tilde B}({\bf x}) {\rm x} (\nabla {\rm x} {\bf \tilde  B}({\bf x}))] \right ) \equiv v_{\rm b}^{\rm 0}({\bf k})\eta
\end{equation}
Here $\rho_{m0}$ is the matter density (baryons and the cold  dark matter) 
at the present epoch.  The compressional velocity field, unlike the vortical
 mode, has a growing mode.
Also unlike the vortical mode (Eq.~(\ref{vort})), the compressional mode of 
 baryonic velocity  couples to the dark matter (\cite{gopal}, \cite{sethi1}).

In Figure~6 we show the angular power spectrum of the secondary temperature
anisotropies generated by the compressional velocity mode. The signal 
has a peak at roughly the angular scale that corresponds to 
 the width of the visibility 
function during 
reionization (for detailed discussion see e.g. \cite{dodelson}). The amplitude 
of this secondary anisotropy is several orders of magnitude smaller
than the observed temperature anisotropies and it is unlikely that this
signal could be detected.

\section{Detectability}
It follows from Figure~1~to~5, that the most important  signal
at small multipoles arises  from tensor polarization anisotropies.
 In particular,
the yet-undetected $B$-mode signal holds the promise of unravelling 
the presence of primordial magnetic fields, as also
 noted by other authors (e.g. \cite{lewis})
In Figure~4, we show
the expected errors on the detection of polarization signal from
the future CMBR mission, Planck surveyor. The expected 1$\sigma$ error, valid for
$\ell \la 100$, is (e.g. \cite{zaldarriaga1}, \cite{prunet}):
\begin{equation} 
\Delta C_\ell = \left ( {2 \over (2 \ell +1)f_{\rm sky}} \right )\left (C_\ell + w^{-1} \right )
\label{erro}
\end{equation}
For Planck surveyor, $f_{\rm sky} \simeq 1$ and $w \simeq  1.7 \times 10^{16}$ for 
one-year integration.  In Figure~4, we 
use the primordial tensor $B$-mode signal for calculating
the expected 1$\sigma$ error from Eq.~(\ref{erro}).  Figure~4 shows  that 
the signal from magnetic fields with strength $\ga  3 \times 10^{-9} \, \rm G$ is  detectable by this future mission. However,  it is likely 
that, except for the $B$ mode signal,  the magnetic field signal will
be buried in a larger signal. However, owing to the non-Gaussianity 
of the  magnetic field signal it might   still be  possible to extract 
this component of the signal (e.g. \cite{lewis}). 

In Figure~5 we show the   TE cross-correlation signal from tensor modes
along with the expected signal from primordial scalar modes with 
$\tau_{\rm reion} = 0.17$, which is in good agreement with the WMAP 
data  of TE cross-correlation \cite{kogut}.
It could be asked if the TE cross-correlation observed by WMAP 
for $\ell \la 100$ could 
be explained as  the tensor signal. 
From Figure~5 it is seen  that the tensor signal at small multipoles is 
roughly a factor of $5$ smaller than the scalar signal. 
And  therefore, as the power spectrum from tangled magnetic 
fields $\propto  B_0^4$, 
 much of the enhancement observed
in the TE cross-correlation  for $\ell \la 10$ 
could   be explainable in terms of the 
tensor signal from primordial magnetic field for $B_0 \simeq 4.5 \times 10^{-9} \, \rm G$. We quantify this notion by
computing  the $\chi^2$ for $\ell \le 15$ for both the best fit 
model from WMAP  and  the tensor  model with $B_0 \simeq 4.5 \times 10^{-9} \, \rm G$ against the detected WMAP signal \footnote{for  details 
of WMAP  data products {\tt http://map.gsfc.nasa.gov}} \cite{kogut}; 
the $\chi^2$ per degree 
of freedom in the two cases is $\simeq 1.7$ and $\simeq 1.8$, respectively. 
Therefore the enhancement can entirely be interpreted in terms of the 
secondary signal from primordial magnetic fields.

 A more realistic possibility  is that
 both primordial 
scalar and tensor modes
gave comparable contribution to the observed signal.  As the 
strength of both these signals for $\ell \la 15$  is roughly $\propto \tau_{\rm reion}^2$ (for details of secondary scalar signal see e.g. \cite{zaldarriaga}), and 
assuming that there is roughly equal contribution from both, the inferred 
value of $\tau_{\rm reion}$ from the analysis of the signal could be smaller
by a factor of  $\sqrt{2}$. To quantify this statement, we did a $\chi^2$ test
to estimate $\tau_{\rm reion}$ by 
adding    the tensor signal with $B_0 \simeq 4.5 \times 10^{-9} \, \rm G$ 
and the primordial scalar signal  with the best-fit cosmological parameters
 from WMAP. 
From this analysis  we obtain 
$\tau_{\rm reion} \simeq 0.11 \pm 0.02$  (1$\sigma$) with $\sigma$ determined
by $\delta\chi^2 = 1$.  A possible  test of 
this hypothesis is  non-gaussianity of the signal at small
multipoles, as the magnetic-field-sourced tensor signal is  not 
Gaussian. 

The tensor signal 
(primary plus secondary) could be appreciable for $\ell \la 100$. In the 
range  $15 \la \ell \la 100$, the tensor and primordial 
scalar signals are  nearly independent of the 
value of $\tau_{\rm reion}$.
While the primordial scalar TE signal anti-correlates for $\ell \ga 40$, 
the tensor signal shows positive cross-correlation in the range $\ell \la 100$,
as seen in Figure~5. The present WMAP data  shows tentative 
detection of TE  anti-correlation  for $\ell \la 100$ \cite{peiris}.
 From $\chi^2$ analysis 
in the range $15 \la \ell \la 100$, we notice that the tensor signal alone 
is a poor fit to the data ($\chi^2$ per degree of freedom of $\simeq 2.1$ as 
opposed to a value of $1.6$ for the primordial scalar model). However 
a sum of these two signals with $B_0 \simeq 4.5 \times 10^{-9} \, \rm G$ is 
a reasonable fit, as it is dominated by the primordial scalar signal. 

 It should be noted 
that for $B_0 \simeq 4.5 \times 10^{-9} \, \rm G$, the tensor temperature 
signal is comparable to  the primordial scalar signal (Figure~3). 
 WMAP analysis obtained an upper 
limit of $\simeq 0.7$  on the ratio of tensor to 
scalar signal (\cite{spergel}). While  this limit is rather weak, 
a more detailed analysis of the 
temperature signal including the effect of tensor mode signal sourced by 
primordial magnetic fields might give independent  constraints 
on the strength of  primordial magnetic fields.

In our $\chi^2$ analysis we use only the diagonal components of 
the Fisher matrix. However,  owing to 
incomplete sky, 
  the signal is correlated, especially for small multipoles,  
 across  neighboring  multipoles. However, a  more comprehensive
analysis taking into this correlation is likely to yield similar conclusions
for the reasons stated above.

Our conclusions are not too sensitive to the value of small scale cut-off
$k_{\rm max}$ or the scale of the filter $k_c$ used to define the normalization
(Eq.~(\ref{normal})) for magnetic field power spectrum index 
 $n = -2.9$  we use throughout the paper. For 
$k_{\rm max} = k_c = 0.05 \, \rm Mpc^{-1}$, the  foregoing 
discussion related to 
tensor mode anisotropies  would be valid for $B_0 \simeq 5 \times 10^{-9} \, \rm G$.  Therefore, the results for TE cross-correlation from 
tensor perturbations can be interpreted 
to put bounds on magnetic fields for only large 
scales $k \la 0.05 \, \rm Mpc^{-1}$. 

The strongest bound on primordial magnetic fields arises from tensor 
perturbations in the pre-recombination era \cite{caprini}. These bounds are
weakest for nearly scale invariant ($n \simeq -3$) magnetic fields 
power spectrum (Eq.~(33) of 
\cite{caprini}) and largely motivated the  choice of the  power spectral
index  we 
consider here. For $n = -2.9$, the bound obtained  by \cite{caprini} is
considerably  weaker
than $B_0 \simeq 4.5 \times 10^{-9} \, \rm G$, the values of interest to us
in this paper. Vector modes might leave observable signature
in the temperature and polarization signal for  $\ell \ga 2000$; the current 
observations give weak bound of  $B_0 \la 8 \times 10^{-9} \, \rm G$ \cite{lewis}. Tangled magnetic-field-sourced primary scalar temperature signal gives even
weaker bounds \cite{koh}.  More recently, Chen et al. \cite{chen} obtained,
from WMAP data analysis, 
a limit of $\la 10^{-8} \, \rm G$ on the primordial magnetic field
strength  for nearly scale invariant spectra we consider here; \cite{chen}
consider  vector mode temperature signal in their analysis and study
possible non-Gaussianity in the WMAP data. 
Another strong constraint on large scale tangled
magnetic fields comes from Faraday rotation of high redshift radio sources (see 
e.g  \cite{widrow}); this constraint is also weaker than the value of 
magnetic field required to explain the enhancement of the TE cross-correlation
signal as seen by WMAP \cite{sethi}. Therefore, the value of $B_0$ required
to give  appreciable contribution 
 to the TE signal is well within the upper limits
on $B_0$ from other considerations.

It should be noted that the entire foregoing discussion on the 
tangled-magnetic-field tensor signal can be mapped to primordial 
tensor modes. The reason for this assertion is that magnetic fields
source tensor modes only prior to the epoch of neutrino decoupling,
and the subsequent  evolution is source free, which is similar to 
the primordial tensor modes which are generated only during the inflationary
epoch and evolve without sources at subsequent times. Therefore, 
an analysis similar to ours could be used to put constraints on the 
relative strength of the tensor to scalar mode contribution 
(for a fixed scale)  and the tensor spectral index of the 
primordial modes.  The main  observational 
difference between such an  interpretation
 and the 
one give  here is that tensor signal sourced by magnetic fields
will not obey Gaussian statistics as opposed to the primordial tensor modes.

\section{Summary and conclusions}
We have computed the secondary anisotropies from the reionization of the 
universe in the presence of tangled primordial magnetic fields. 
Throughout our analysis we use the nearly scale invariant magnetic field
power spectrum with $n = -2.9$. 
For vector modes, we compute the secondary temperature 
 and $E$ and $B$ mode
polarization auto-correlation signal. For scalar modes,  the 
results for secondary
temperature angular power spectrum from compressional velocity modes
are  presented. For tensor modes, in addition to the 
secondary temperature and polarization angular power spectra, we 
compute the TE cross-correlation signal and compare it with the 
existing WMAP data; we also recompute the primary signal for 
tensor modes. Whenever possible we compare our results with the results 
existing in the literature. In particular, Lewis \cite{lewis} recently
computed fully-numerically the vector and tensor primary and 
secondary  temperature and  polarization power spectra. We compare our 
semi-analytic results with this analysis and find good agreement. Seshadri and
Subramanian  \cite{seshadri1} computed the secondary temperature anisotropies from vector modes.
Our results are in good agreement with their conclusion. Mack et al. \cite{mack} 
computed primary signal from vector and tensor modes using the formalism
we adopt in this paper. Our results are in disagreement with their results 
for $\ell \ga 75$, and we have given  reasons for our 
disagreement in the discussion above.  In addition to comparison with
existing literature, we
also give new results for secondary 
TE cross-correlation from tensor modes  and secondary 
temperature angular power spectrum from scalar modes. 

We discuss below the details of 
 expected signal from each of the perturbation mode:

 {\it Vector modes}: The secondary temperature and polarization 
signals from the vector modes is shown are  Figure~1~and~2. The 
secondary temperature signal increases $\propto \ell^{2.4}$ for 
$\ell \ga 50$ and reaches
a value $\simeq 0.1 \, \rm (\mu k)^2$ for $\ell \simeq 10^4$,
in agreement with the analysis of \cite{seshadri1}. For small $\ell$ the signal
is very small ($\la 10^{-4}\, \rm (\mu k)^2$) and for  large
$\ell$ the secondary signal is  smaller than the primary signal 
(e.g. \cite{lewis}) and 
therefore it is unlikely  that the signature of reionization could be detected
in the vector-mode  temperature anisotropies. The polarization signal, shown
in Figure~2, is sourced  by the free-streaming  of dipole at the epoch of 
recombination. This signal dominates the primary signal for $\ell \la 10$, but 
is several orders of magnitude smaller than the expected signal from 
tensor modes. 

{\it Scalar modes}: We only compute the secondary temperature 
anisotropies from compressional velocity modes in this case. As seen in
Figure~6, this contribution is several orders of magnitude smaller than
the already-detected primary signal and therefore its effects are 
unlikely to be detectable. 

{\it Tensor modes}: As seen from Figures~4 and~5, the most interesting CMBR
anisotropy signal for $\ell \la 100$ is from these modes. 
The secondary $B$-mode signal from tensor 
modes is detectable by future CMBR mission Planck surveyor for $B_0 \simeq 3\times 10^{-9} \, \rm G$ . The 
tensor TE cross
correlation from primordial magnetic fields can explain the observed
enhancement of the observed signal for $\ell \la 10$ 
 by WMAP for $B_0 \simeq 4.5 \times 10^{-9} \, \rm G$ if the 
primordial magnetic fields are generated during the 
epoch of inflation. Assuming that tensor modes make a significant 
contribution to the observed enhancement, the bounds on the optical
depth to the surface of reionization, $\tau_{\rm reion}$ are weaker 
by roughly a factor of $\sqrt{2}$. This hypothesis can be borne/ruled  out
by testing the Gaussianity of the signal for $\ell \la 10$

\section*{Acknowledgment}
We would like to thank A. Lewis for prompt reply to our queries and 
to T. R. Seshadri   and K. Subramanian for useful discussion.

\section*{Appendix A}
In this section, we briefly discuss the terminology and present the 
complete expressions for the vector and tensor power spectra $\Pi^{V}(k)$ and
 $\Pi^{T}(k)$ (\cite{mack}).
The energy momentum tensor for magnetic fields for a single Fourier mode is 
 a convolution of different Fourier  modes and is given by:
\begin{equation}
T_{ij}({\mathbf{k}})=\int d^{3}q \left[ \tilde B_{i}({\mathbf{q}})\tilde B_{j}({\mathbf{k-q}})-\frac{1}{2} \delta_{ij}\tilde B_{m}({\mathbf{q}})\tilde B_{m}({\mathbf{k-q}})\right]
\end{equation}
The energy-momentum tensor has non-vanishing scalar, vector, and tensor 
components. The vector and tensor components, in Fourier space, 
 are defined as:
\begin{eqnarray}
\Pi_{i}^{V}=P_{ip}\hat{k}_{q}T_{pq} \label{vecso} \\
\Pi_{ij}^{T}=\left(P_{ip}P_{jq}-\frac{1}{2}P_{ij}P_{pq}\right)T_{pq}
\end{eqnarray}
 Here $P_{ij} = \delta_{ij}  - \hat k_i \hat k_j$.
  The vector and tensor anisotropic stress are then defined as the two-point 
correlations of the above components as:
\begin{eqnarray}
\langle \Pi_{i}^{V}({\mathbf{k}})\Pi_{i}^{V}({\mathbf{k^{'}}}) \rangle 
\equiv 2|\Pi^{(V)}(k)|^{2}\delta({\mathbf{k+k^{'}}})  \\
\langle \Pi_{ij}^{T}({\mathbf{k}})\Pi_{ij}^{T}({\mathbf{k^{'}}}) \rangle \equiv 4|\Pi^{(T)}(k)|^{2}\delta({\mathbf{k+k^{'}}}) \label{powspec_b}
\end{eqnarray}
 By evaluating the above correlations as also given in \cite{mack}, we 
can arrive at 
the following approximate expression for the power spectra for $n<-3/2$.
\begin{eqnarray}
|\Pi_{V}(k)|^{2}=\frac{A^{2}}{64\pi^{4}(n+3)}k^{2n+3}\\
|\Pi_{T}(k)|^{2}=\frac{2A^{2}}{64\pi^{4}(n+3)}k^{2n+3}
\end{eqnarray}
Here, $A$ is the normalization of the magnetic power spectrum given in Eq.~(\ref{normal})

\section*{Appendix B}
Gravitational waves correspond to transverse,traceless perturbations to the 
metric: $\delta g_{ij}=2a^{2}(\eta)h_{ij}$ with $h_{ii}=\hat{k_{i}}h_{ij}=0$. 
Since $h_{ij}$ is a stochastic variable we can define its power spectrum as:
\begin{equation}
\langle h_{ij}({\mathbf{k}},\eta)h_{ij}({\mathbf{k^{'}}},\eta) \rangle =4|h(k,\eta)|^{2}\delta({\mathbf{k+k^{'}}})
\end{equation}
 The evolution of $h_{ij}$ then follows from the tensor Einstein 
equation (see e.g. \cite{hu4}) ,
\begin{equation}
\ddot{h}+2\frac{\dot{a}}{a}\dot{h}+k^{2}h=8\pi G S(k,\eta)
\label{ten_evol}
\end{equation}
The source on the RHS is the tensor anisotropic stress of the plasma which 
is defined as: $S(k,\eta) = \Pi^T(k)/a^2$  (Eq.~(\ref{powspec_b})). 
 We assume that 
the primordial magnetic fields are generated by some mechanism at a very
 early epoch $\eta_{in}$. It  was recently shown by Lewis \cite{lewis} that 
after the neutrino decoupling epoch $\eta_\star$ the 
neutrino start free-streaming and develop significant anisotropic
 stress which cancel the  anisotropic stress of the
 primordial magnetic fields to the leading order for super-horizon modes, 
 resulting in negligible net 
anisotropic 
stress in the plasma.  We can thus assume that 
for $\eta \gg \eta_\star$, $S({\mathbf{k}},\eta)=0$ and for $\eta \ll \eta_\star$, 
$S({\mathbf{k}},\eta)=\Pi^{T}(k)/a^{2}$ where $\Pi^{T}(k)$ is the
 magnetic tensor anisotropic stress as 
defined in Eq.~(\ref{powspec_b}). We now derive the solutions to 
Eq.~(\ref{ten_evol}) in various regimes.  
The evolution of the scale factor $a(\eta)$ is given by the Friedmann equation:
\begin{equation}
{\dot{a}}^{2}=H_{0}^{2}(\Omega_{m} a +\Omega_{\gamma}+\Omega_{\nu}+\Omega_{\Lambda} a^{4})
\end{equation} 
Here, $\Omega_{m,\gamma,\nu,\Lambda}$ are the fractional densities in matter,
radiation,neutrinos and cosmological constant respectively. 
 Approximate solutions in the radiation-dominated and
 matter-dominated epoch are $a(\eta)=2\sqrt{\frac{\Omega_{\gamma}+\Omega_{\nu}}{\Omega_{m}}}\frac{\eta}{\eta_{0}}$ and $a(\eta)=\left(\frac{\eta}{\eta_{0}}\right)^{2}$ respectively. 
Using the above form for the scale-factor we can rewrite Eq.~(\ref{ten_evol})  for $ \eta_{in}< \eta < \eta_{\star} $ as:
\begin{equation}
\ddot{h}+\frac{2}{\eta}\dot{h}+k^{2}h=\frac{3R_{\gamma}\Pi^{T}(k)}{\rho_{\gamma}}\frac{1}{\eta^{2}}
\label{ten_evol1}
\end{equation}
Here, $R_{\gamma}=\Omega_{\gamma}/(\Omega_{\gamma}+\Omega_{\nu}) \simeq 0.6$.
 $\rho_{\gamma}$ is the 
CMBR  energy density. Eq.~(\ref{ten_evol1})
 can be solved exactly using the Green's 
function technique to give \cite{mack}: 
\begin{equation}
h(k,\eta)=\frac{3R_{\gamma}\Pi^{T}(k)}{\rho_{\gamma}}\int_{\eta_{in}}^{\eta}d\eta^{'}\frac{\sin[k(\eta-\eta^{'})]}{\eta^{'}}
\end{equation}
For super-horizon modes $k\eta\ll 1$, the above form can be simplified to give:
\begin{equation}
h(k,\eta)\approx\frac{3R_{\gamma}\Pi^{T}(k)}{\rho_{\gamma}}\int_{\eta_{in}}^{\eta}d\eta^{'}\frac{k(\eta-\eta^{'})}{k\eta\eta^{'}}=\frac{3R_{\gamma}\Pi^{T}(k)}{\rho_{\gamma}}\ln\left(\frac{\eta}{\eta_{\rm in}}\right)
\end{equation}
For $\eta \gg \eta_\star$, the evolution of $h$ is given by the homogeneous
 solutions in the radiation and matter-dominated regimes:
\begin{eqnarray}
h_{rad}(k,\eta)=A_{1}j_{0}(k\eta)\\
h_{mat}(k,\eta)=A_{2}\frac{j_{1}(k\eta)}{k\eta}
\end{eqnarray}
The coefficients $A_{1}$ and $A_{2}$ are determined by matching the
 super-horizon solutions at the two transitions $\eta_{\star}$ and $\eta_{eq}$. We thus get
\begin{equation}
A_{2}=3A_{1}=\frac{9R_{\gamma}\Pi^{T}(k)}{\rho_{\gamma}}\ln\left(\frac{\eta_{\star}}{\eta_{\rm in}}\right)
\end{equation}
Thus, the full expression for the matter-dominated solution can be written as:
\begin{equation}
\dot{h}_{mat}(\eta,k)=\frac{9R_{\gamma}\Pi^{T}(k)}{\rho_{\gamma}}\ln\left(\frac{\eta_{\star}}{\eta_{\rm in}}\right)\frac{j_{2}(k\eta)}{\eta}
\label{finsol}
\end{equation}
This solution  is used for solving tensor temperature and polarization
primary and secondary anisotropies. 
 Few assumptions have been made in deriving the above expression. 
Firstly, the transition between radiation dominated to matter-dominated 
region has been assumed to be instantaneous. This however does not affect 
the evolution of modes with wave-length greater than the width of 
transition $k\eta_{\rm eq} \la 1$. Moreover, only super-horizon solutions have
 been used to match the solutions for $h$ at different transitions. These 
simplifications however do not affect the results quotes for small
multipoles as discussed in the main section.

\subsection*{Tight-coupling tensor quadrupole}
 In the tight-coupling regime, $z \ga 1100$, 
 to lowest order in mean-free path, we have
 $P^{T}=-\dot{h}/(3\dot{\tau})$ \cite{mack}. We however use the expression
 accurate to the second
 order in mean-free path as is done for the scalar modes in
 \cite{zaldarriaga2}.
 Using the Boltzmann equation for the evolution of tensor modes we get
 the following
 equation for $P^{T}(\mathbf{k},\eta)$ in the tight-coupling  limit:
 \begin{equation}
 \dot{P}+\frac{3}{10}\dot{\tau}P=-\frac{\dot{h}}{10}
 \end{equation}
The lowest order solution to this equation is obtained by neglecting
the $\dot P$ in the equation, which gives, $P = -\dot h/(3 \dot\tau)$. The 
above equation however  can be solved exactly  to give:
 \begin{equation}
 P(\eta)=\int_{0}^{\eta}d\eta{'}\dot{h}e^{-\frac{3}{10}[\tau(\eta^{'})-\tau(\eta)]}
\label{tenquadr}
 \end{equation}
We use the standard recombination history for computing $\tau$.

\newpage
\begin{figure}
\epsfig{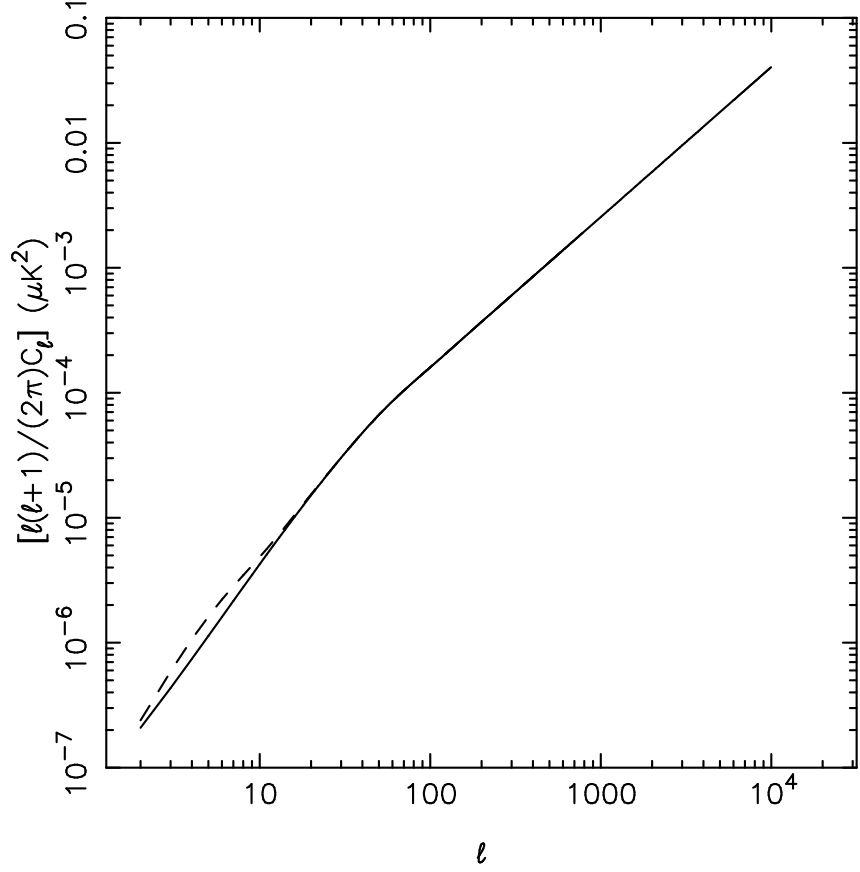}
\caption{The secondary 
temperature angular power spectrum from vector modes  is shown. The
solid and the dashed lines correspond to the contribution from 
vorticity and the total signal, respectively (see text for details).
The power spectrum is plotted for $B_0 = 3 \times 10^{-9}$ and $n = -2.9$
(Eq.~(\ref{normal})).  }
\label{fig:f1}
\end{figure}
\newpage
\begin{figure}
\epsfig{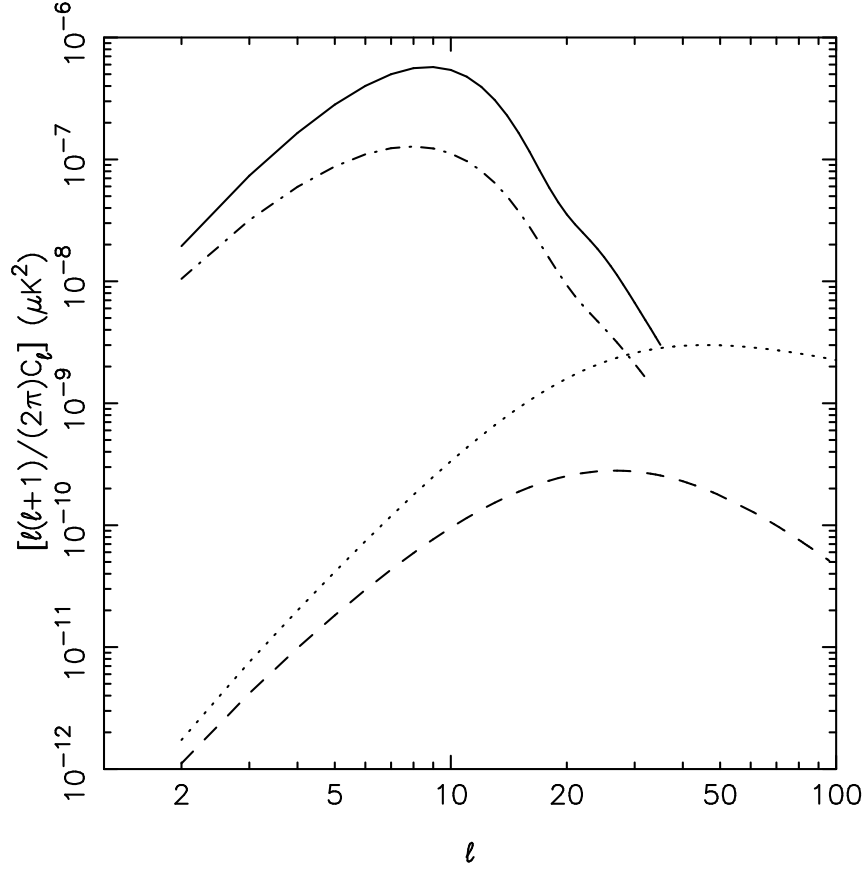}
\caption{The secondary 
polarization  angular power spectrum from vector modes is shown. The
solid and the dot-dashed lines correspond, respectively, 
 to  the $B$ and $E$ mode
contribution from  the free-streaming quadrupole (Eq.~(\ref{eq:quadr})). 
The dotted and dashed curves $B$ and $E$ mode signals that arises from
 the source term given by   Eq.~(\ref{eq:quadr1}). The power 
spectra are plotted for  $B_0 = 3 \times 10^{-9}$ and $n = -2.9$
(Eq.~(\ref{normal})). }
\label{fig:f2}
\end{figure}

\newpage
\begin{figure}
\epsfig{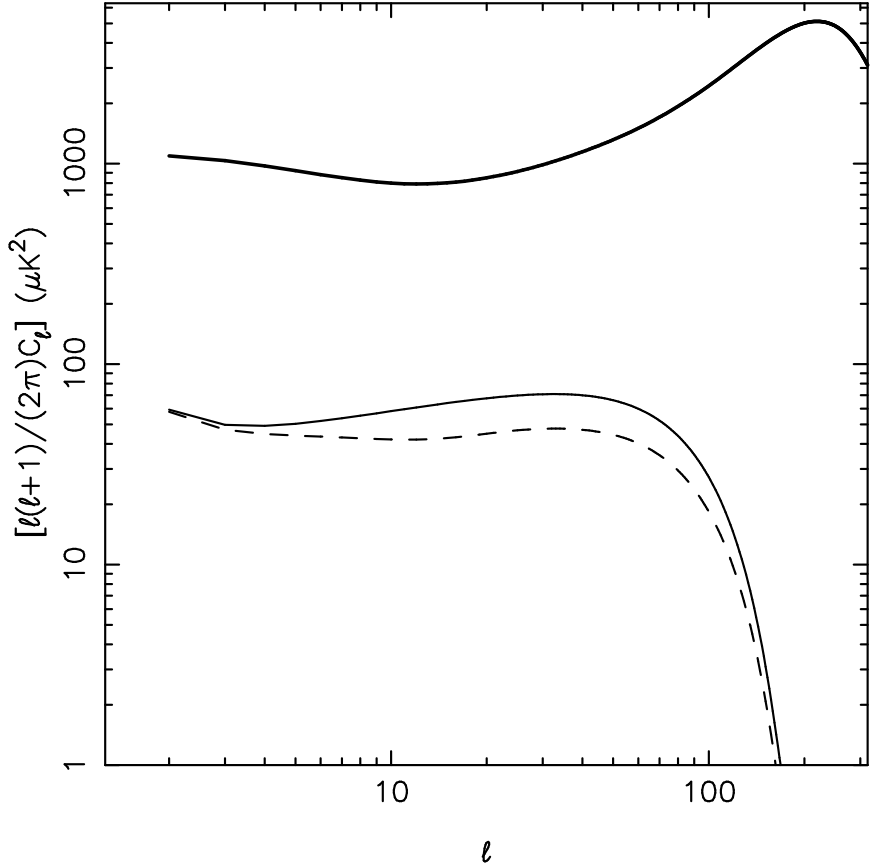}
\caption{The contribution of tensor modes to the temperature power
spectrum is shown. The solid and dashed lines give, respectively,
the power spectra  without and with reionization. The thick
solid line, shown here for comparison, 
correspond to  the temperature power spectrum from scalar modes, 
for the best-fit parameters from WMAP (\cite{spergel}). The power spectra are
 plotted for $B_0 = 3 \times 10^{-9}$,  $n = -2.9$
(Eq.(\ref{normal})), and $\eta_\star/\eta_{\rm in} = 10^{18}$ (Eq.~(\ref{finsol})). }
\label{fig:f3}
\end{figure}

\newpage
\begin{figure}
\epsfig{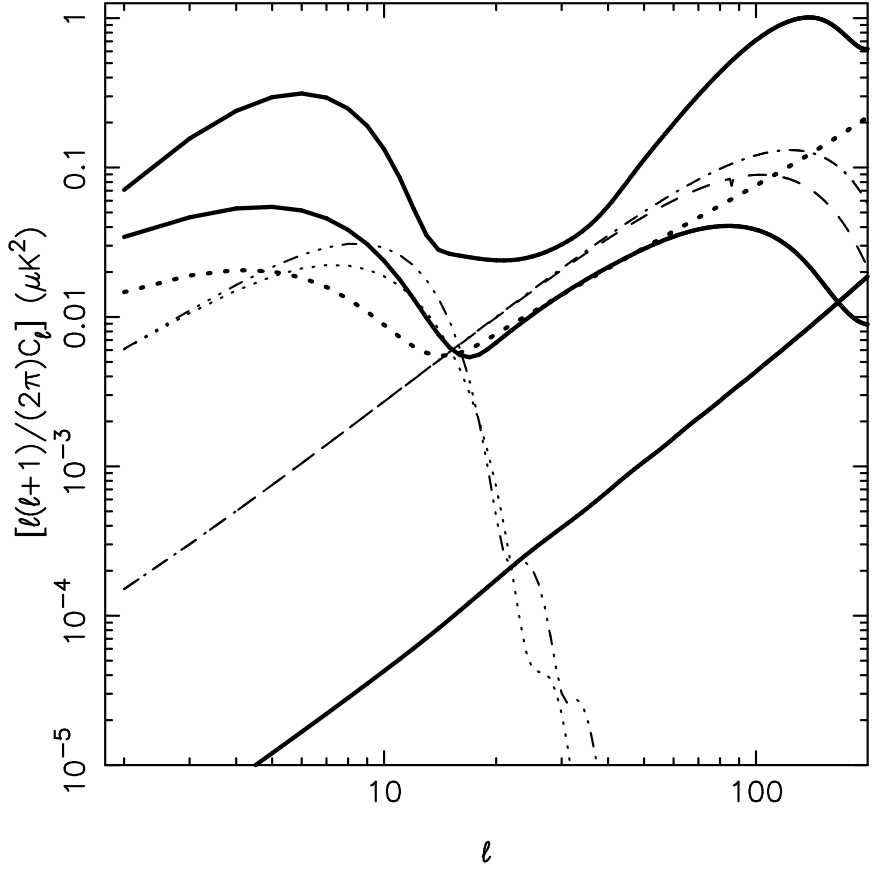}
\caption{The tensor secondary and primary    polarization  power spectra 
are shown along with the expected signals from primordial scalar
and tensor modes. The dot-dot-dot dashed and thin dotted 
lines correspond to the secondary $E$ and $B$ mode power spectra, 
respectively. The dot-dashed and dashed lines give the primary $E$ and 
$B$ mode power spectra. The two top
solid lines, shown here for comparison, 
correspond to  the  $E$ and $B$ mode  power spectra from  
primordial scalar modes, for the best-fit parameters from WMAP (\cite{spergel}).  For $B$ mode signal we assume the ratio of tensor to scalar quadrupole $T/S = 0.7$ and the tensor spectral index $n_t = 0$. The bottom solid lines
shows the $B$-mode signal expected from gravitational lensing. The thick dashed
line shows the 1-$\sigma$ errors expected from the future CMBR experiment 
Planck surveyor for one year of integration (Eq.~(\ref{erro})). 
The power spectra are
 plotted for $B_0 = 3 \times 10^{-9}$, $n = -2.9$ 
(Eq.(\ref{normal})), and $\eta_\star/\eta_{\rm in} = 10^{18}$ (Eq.~(\ref{finsol})).  }
\label{fig:f4}
\end{figure}

\newpage
\begin{figure}
\epsfig{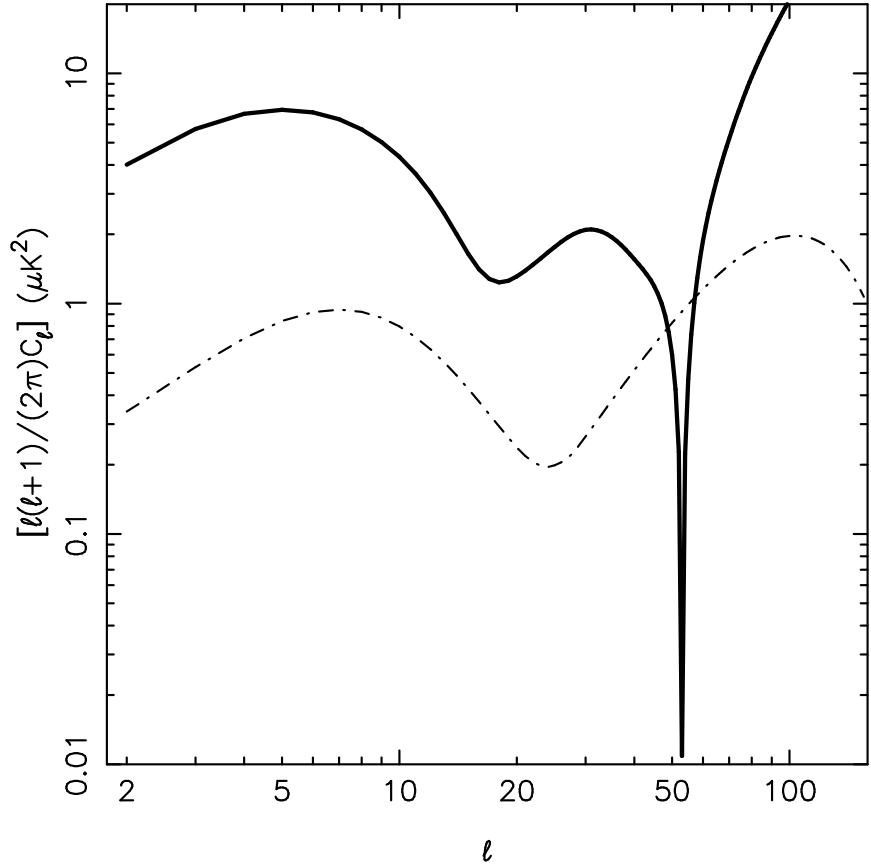}
\caption{The tensor TE cross-correlation  power spectra 
are shown along with the expected signal from primordial scalar  modes.
 The dot-dashed shows the TE cross-correlation (secondary plus 
primary) from tangled magnetic 
fields.  The thick 
solid line, shown here for comparison, 
correspond to  the (absolute value of) 
  TE cross-correlation  power spectrum from  
primordial scalar modes, for the best-fit parameters from WMAP (\cite{spergel}).  The power spectrum is 
 plotted for $B_0 = 3 \times 10^{-9}$, $n = -2.9$
(Eq.(\ref{normal})),  and $\eta_\star/\eta_{\rm in} = 10^{18}$ (Eq.~(\ref{finsol})). }
\label{fig:f5}
\end{figure}

\newpage
\begin{figure}
\epsfig{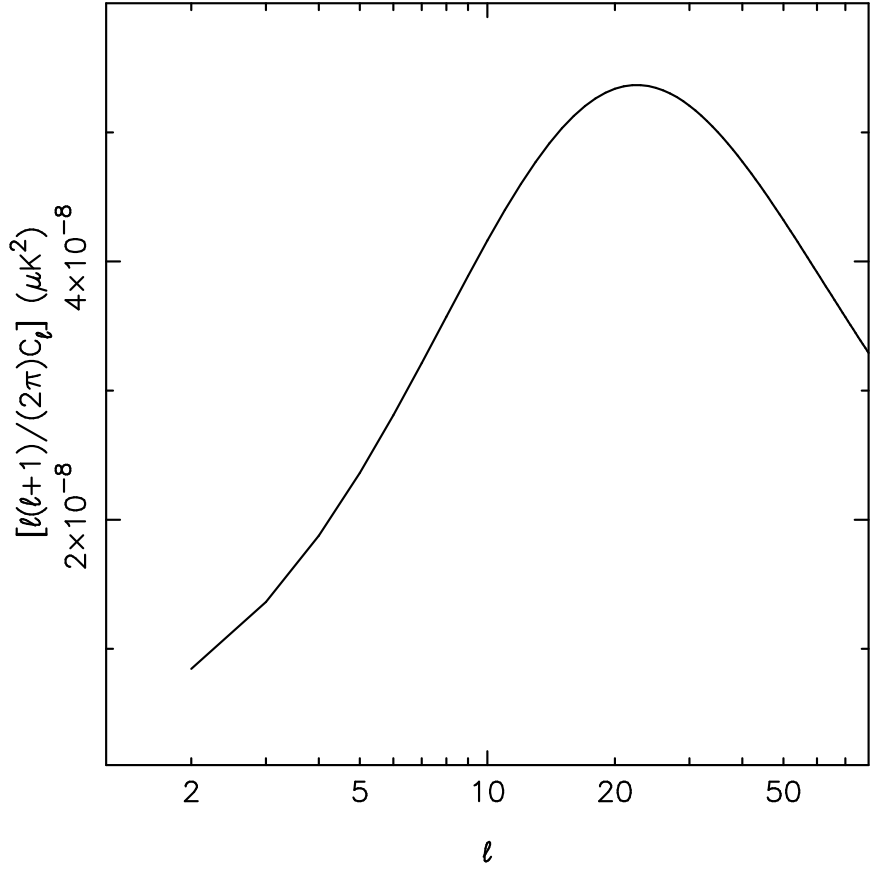}
\caption{The secondary temperature power spectrum  from 
scalar mode perturbations, seeded
by tangled  magnetic fields, is shown.  The power spectrum is 
 plotted for $B_0 = 3 \times 10^{-9}$ and $n = -2.9$
(Eq.(\ref{normal})). }
\label{fig:f6}
\end{figure}

\end{document}